\documentclass{aastex701}

\usepackage{graphicx}
\usepackage{xcolor}

\begin{document} 

\title{Generating LSB-optimised synthetic images for simulated galaxies}

\author[orcid=0000-0002-3930-2757]{Maarten Baes}
\affiliation{Sterrenkundig Observatorium, Universiteit Gent, Krijgslaan 299, B-9000 Gent, Belgium}
\email{maarten.baes@ugent.be}

\author[orcid=0000-0002-4479-4119]{Peter Camps}
\affiliation{Sterrenkundig Observatorium, Universiteit Gent, Krijgslaan 299, B-9000 Gent, Belgium}
\email{peter.camps@ugent.be}

\author[orcid=0000-0002-0206-8231]{Andrea Gebek}
\affiliation{Sterrenkundig Observatorium, Universiteit Gent, Krijgslaan 299, B-9000 Gent, Belgium}
\email{andrea.gebek@ugent.be}

\author[orcid=0009-0003-9692-9382]{Arno Lauwers}
\affiliation{Sterrenkundig Observatorium, Universiteit Gent, Krijgslaan 299, B-9000 Gent, Belgium}
\email{arno.lauwers@ugent.be}

\author[orcid=0000-0002-0668-5560]{Joop Schaye}
\affiliation{Leiden Observatory, Leiden University, PO Box 9513, 2300 RA Leiden, The Netherlands}
\email{schaye@strw.leidenuniv.nl}

\author[orcid=0000-0003-3665-4519]{Paul Vauterin}
\affiliation{Sterrenkundig Observatorium, Universiteit Gent, Krijgslaan 299, B-9000 Gent, Belgium}
\email{paul.vauterin@ugent.be}


\begin{abstract}
\noindent
We introduce an emission–biasing scheme in the {\tt SKIRT} radiative transfer code that enables efficient generation of synthetic galaxy images optimized for low-surface-brightness (LSB) science. Standard Monte Carlo radiative transfer simulations achieve high signal-to-noise in bright regions but require prohibitively many photon packets to reach reliable depth in galaxy outskirts. By assigning stellar particles bias factors that scale with their smoothing lengths, our method boosts photon emission from low-density regions while conserving energy through weight corrections. Tests on a Milky-Way-like galaxy from the TNG50 cosmological simulation show that bias factors proportional to the smoothing length substantially extend the reliable LSB regime, providing an inexpensive improvement for deep synthetic imaging of simulated galaxies.
\end{abstract}

\keywords{radiative transfer --- dust: extinction --- galaxies: structure}


\section{Introduction} 
\label{sec:introduction}

\noindent
Cosmological hydrodynamical simulations are essential tools for studying galaxy evolution. Synthetic observations bridge the gap between simulations and real data by allowing simulated galaxies to be analysed within the same observational framework as their observed counterparts. The {\tt SKIRT} Monte Carlo radiative transfer code \citep{Baes2011b, Camps2015a, Camps2020} is a widely used tool for generating realistic dust-aware synthetic data products across a broad range of cosmological simulations \citep[e.g.,][]{Trayford2017, Camps2018a, RodriguezGomez2019, Baes2024a, Bottrell2024}.

Low-surface-brightness (LSB) structures, such as stellar halos, tidal streams, and diffuse outer disks, offer powerful constraints on galaxy assembly histories and dark matter distributions. However, synthetic images produced with standard Monte Carlo radiative transfer are typically not optimised for LSB science: photon packets naturally concentrate in high-surface-brightness (HSB) regions, yielding excellent signal-to-noise (SNR) there but poor performance in faint outskirts. Achieving high SNR in LSB regions requires launching extremely large numbers of photon packets, resulting in prohibitively long runtimes.

In this Research Note, we describe a new emission-biasing feature in {\tt SKIRT} designed to improve LSB sensitivity without increasing computational cost.


\section{The generation of LSB-optimised images}

\begin{figure*}
\includegraphics[width=\textwidth]{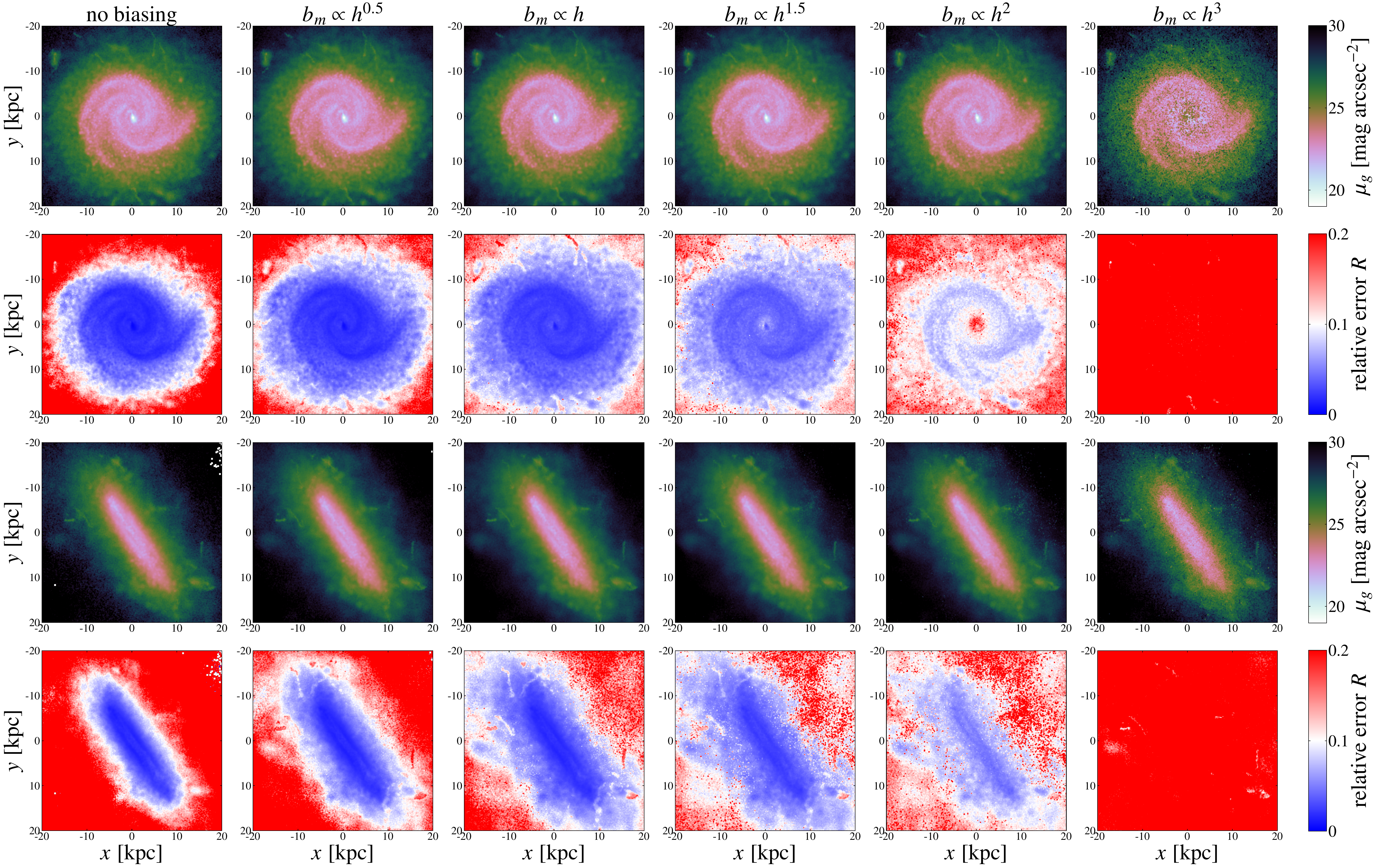}
\caption{$g$-band images and relative error maps for {\tt SKIRT} simulations of the galaxy TNG000008. Images are shown for face-on and edge-on orientations (rows) and six emission-bias schemes, ranging from no biasing to $b_m \propto h^3$ (columns).}
\label{imagesh.fig}
\end{figure*}

\noindent
In a Monte Carlo radiative transfer simulation, a large number of photon packets are launched stochastically from a source. In the case of simulated galaxies, the source consists of a set of stellar particles. After emission, each photon packet propagates through the dusty interstellar medium, may undergo interactions, and is eventually detected once it exits the system.

The first step is to randomly select a particle $m$ from which a photon packet is launched. As the number of photon packets launched from a particle should be proportional to that particle’s contribution to the galaxy's total luminosity, the appropriate probability density function (pdf) is
\begin{equation}
p(m) = \frac{L_m}{\sum_n L_n} \quad m = 1, \ldots, M.
\label{pdfp}
\end{equation}
A general optimisation technique frequently used in Monte Carlo radiative transfer is biasing: one draws events from a modified pdf and compensates for this by multiplying the photon packet's weight by a correction factor \citep{Steinacker2013}. Assigning a bias factor $b_m$ to each particle yields the biased pdf
\begin{equation}
q(m) = \frac{b_m L_m}{\sum_n b_n L_n}, 
\label{pdfq}
\end{equation}
and the corresponding weight correction factor becomes
\begin{equation}
w(m) = \frac{p(m)}{q(m)} = \frac{\sum_n b_n L_n}{b_m \sum_n L_n}.
\end{equation}
We have implemented this emission-biasing mechanism in {\tt SKIRT}.\footnote{More generally, {\tt SKIRT} employs composite biasing \citep{Baes2016} to prevent excessively large correction factors. For clarity, we present here only the basic formulation.} To enhance LSB sensitivity, the bias factors should preferentially increase photon emission from low-density regions. A natural choice is to scale $b_m$ with a power of the particle’s smoothing length $h$, which is typically set to the distance to the $N$th nearest neighbour. Indeed, because smoothing lengths increase in low-density regions, they provide a natural proxy for identifying particles in the galaxy outskirts where LSB emission originates.


\section{Illustration}

\noindent
We applied this biasing scheme to synthetic images of the simulated galaxy TNG000008, a Milky-Way-like system from the $z=0$ snapshot of the TNG50 simulation \citep{Pillepich2019, Nelson2019b}. The galaxy contains 691k stellar particles, a total stellar mass of $3.73\times10^{10}~{\text{M}}_\odot$, and ${\text{SFR}} = 3.98~{\text{M}}_\odot~{\text{yr}}^{-1}$. Using the dust-insertion recipe outlined in \citet{Baes2024a}, the galaxy has a dust mass of $5.59\times10^7~{\text{M}}_\odot$. We assign each particle a smoothing length corresponding to the distance to its 32nd-nearest neighbour, as is conventionally adopted \citep[e.g.,][]{Torrey2015, RodriguezGomez2019, Schulz2020, Baes2024a}.

The first and third panels of the left column of Fig.~\ref{imagesh.fig} show face-on and edge-on $g$-band images generated using $10^8$ photon packets with natural weighting (i.e., no emission biasing). Each image spans 40~kpc and contains $200\times200$ pixels. The second and fourth panels display the corresponding $R$ maps, a reliability statistic calculated according to the methodology of \citet{X5MonteCarloTeam2003} and \citet{Camps2018b, Camps2020}. Pixel values with $R<0.1$ are deemed reliable, and in this case $R$ can be considered as the relative error on the observed surface brightness. On the other hand, pixels with $0.1<R<0.2$ are questionable, and those with $R>0.2$ are unreliable.  With natural weighting, only the central HSB regions reach the $R<0.1$ threshold.

The remaining columns show results obtained with different emission-bias schemes. Because particles with the largest smoothing lengths reside in the outer parts of the galaxy, biasing increases photon emission in these regions while reducing it in the centre. This redistribution is subtle in the images but clearly evident in the relative error maps. For $b_m \propto h$ and $b_m \propto h^{1.5}$, the reliable region ($R<0.1$) expands significantly, though the reliability in the inner regions decreases modestly. In contrast, $b_m \propto h^2$ shifts too many photon packets to the outskirts, making the central regions unreliable. For even higher powers, such as $b_m \propto h^3$, this effect becomes more extreme and the images are almost completely unreliable. We assessed all biasing schemes by counting pixels with $R<0.1$ or $R<0.2$. The choice $b_m \propto h$ performs best, and this conclusion is robust against changes in viewing angle and photon-packet number.


\section{Conclusions}

\noindent
We conclude that emission biasing provides a simple and effective way to generate LSB-optimised synthetic images of simulated galaxies. The implementation in {\tt SKIRT} assigns a user-defined bias factor $b_m$ to each stellar particle. Choosing $b_m \propto h$ offers an optimal balance: it enhances signal-to-noise in the LSB regime while maintaining sufficient reliability in the bright central regions. This approach should facilitate synthetic comparisons with deep imaging surveys such as MATLAS \citep{Bilek2020}, LIGHTS \citep{Trujillo2021}, or VST-SMASH \citep{Tortora2024}, and forthcoming dedicated missions such as ARRAKIHS \citep{Guzman2024}.


\begin{acknowledgments}
MB and PV acknowledge funding from the Belgian Science Policy Oﬃce (BELSPO) through
the PRODEX project ``ARRAKIHS Science Development at UGent (ASDUG)'' (C4000147090). AL is a PhD Fellow of the Flemish Fund for Scientific Research (FWO-Vlaanderen, grant 1193525N).
\end{acknowledgments}


\bibliography{mybib_nameyear}{}

@PREAMBLE{ {\providecommand{\noopsort}[1]{}} }

@ARTICLE{Baes2011b,
       author = {{Baes}, Maarten and {Verstappen}, Joris and {De Looze}, Ilse and {Fritz}, Jacopo and {Saftly}, Waad and {Vidal P{\'e}rez}, Edgardo and {Stalevski}, Marko and {Valcke}, Sander},
        title = "{Efficient Three-dimensional NLTE Dust Radiative Transfer with SKIRT}",
      journal = {\apjs},
         year = 2011,
        month = oct,
       volume = {196},
        pages = {22},
          doi = {10.1088/0067-0049/196/2/22}
}

@ARTICLE{Baes2016,
       author = {{Baes}, Maarten and {Gordon}, Karl D. and {Lunttila}, Tuomas and {Bianchi}, Simone and {Camps}, Peter and {Juvela}, Mika and {Kuiper}, Rolf},
        title = "{Composite biasing in Monte Carlo radiative transfer}",
      journal = {\aap},
         year = 2016,
        month = may,
       volume = {590},
        pages = {A55},
          doi = {10.1051/0004-6361/201528063}
}

@ARTICLE{Baes2024a,
       author = {{Baes}, Maarten and {Gebek}, Andrea and {Tr{\v{c}}ka}, Ana and {Camps}, Peter and {van der Wel}, Arjen and {Abdurro'uf} and {Andreadis}, Nick and {Tulu}, Sena Bokona and {Emana}, Abdissa Tassama and {Fritz}, Jacopo and {Kelly}, Raymond and {Kova{\v{c}}i{\'c}}, Inja and {La Marca}, Antonio and {Martorano}, Marco and {Mosenkov}, Aleksandr and {Nersesian}, Angelos and {Rodriguez-Gomez}, Vicente and {Tortora}, Crescenzo and {Vander Meulen}, Bert and {Wang}, Lingyu},
        title = "{The TNG50-SKIRT Atlas: Post-processing methodology and first data release}",
      journal = {\aap},
         year = 2024,
        month = mar,
       volume = {683},
        pages = {A181},
          doi = {10.1051/0004-6361/202348418}
}

@ARTICLE{Bilek2020,
       author = {{B{\'\i}lek}, Michal and {Duc}, Pierre-Alain and {Cuillandre}, Jean-Charles and {Gwyn}, Stephen and {Cappellari}, Michele and {Bekaert}, David V. and {Bonfini}, Paolo and {Bitsakis}, Theodoros and {Paudel}, Sanjaya and {Krajnovi{\'c}}, Davor and {Durrell}, Patrick R. and {Marleau}, Francine},
        title = "{Census and classification of low-surface-brightness structures in nearby early-type galaxies from the MATLAS survey}",
      journal = {\mnras},
         year = 2020,
        month = oct,
       volume = {498},
        pages = {2138-2166},
          doi = {10.1093/mnras/staa2248}
}

@ARTICLE{Bottrell2024,
       author = {{Bottrell}, Connor and {Yesuf}, Hassen M. and {Popping}, Gerg{\"o} and {Omori}, Kiyoaki Christopher and {Tang}, Shenli and {Ding}, Xuheng and {Pillepich}, Annalisa and {Nelson}, Dylan and {Eisert}, Lukas and {Gao}, Hua and {Goulding}, Andy D. and {Kalita}, Boris S. and {Luo}, Wentao and {Greene}, Jenny E. and {Shi}, Jingjing and {Silverman}, John D.},
        title = "{IllustrisTNG in the HSC-SSP: image data release and the major role of mini mergers as drivers of asymmetry and star formation}",
      journal = {\mnras},
         year = 2024,
        month = jan,
       volume = {527},
        pages = {6506-6539},
          doi = {10.1093/mnras/stad2971}
}

@ARTICLE{Camps2015a,
       author = {{Camps}, P. and {Baes}, M.},
        title = "{SKIRT: An advanced dust radiative transfer code with a user-friendly architecture}",
      journal = {Astronomy and Computing},
         year = 2015,
        month = mar,
       volume = {9},
        pages = {20-33},
          doi = {10.1016/j.ascom.2014.10.004}
}

@ARTICLE{Camps2018a,
       author = {{Camps}, Peter and {Tr{\v{c}}ka}, Ana and {Trayford}, James and {Baes}, Maarten and {Theuns}, Tom and {Crain}, Robert A. and {McAlpine}, Stuart and {Schaller}, Matthieu and {Schaye}, Joop},
        title = "{Data Release of UV to Submillimeter Broadband Fluxes for Simulated Galaxies from the EAGLE Project}",
      journal = {\apjs},
         year = 2018,
        month = feb,
       volume = {234},
        pages = {20},
          doi = {10.3847/1538-4365/aaa24c}
}

@ARTICLE{Camps2018b,
       author = {{Camps}, Peter and {Baes}, Maarten},
        title = "{The Failure of Monte Carlo Radiative Transfer at Medium to High Optical Depths}",
      journal = {\apj},
         year = 2018,
        month = jul,
       volume = {861},
        pages = {80},
          doi = {10.3847/1538-4357/aac824}
}

@ARTICLE{Camps2020,
       author = {{Camps}, P. and {Baes}, M.},
        title = "{SKIRT 9: Redesigning an advanced dust radiative transfer code to allow kinematics, line transfer and polarization by aligned dust grains}",
      journal = {Astronomy and Computing},
         year = 2020,
        month = apr,
       volume = {31},
        pages = {100381},
          doi = {10.1016/j.ascom.2020.100381}
}

@INPROCEEDINGS{Guzman2024,
       author = {{Guzm{\'a}n}, Rafael},
        title = "{ARRAKIHS: The New ESA F-Class Mission to Investigate the Nature of Dark Matter}",
    booktitle = {EAS2024, European Astronomical Society Annual Meeting},
         year = 2024,
        month = jul,
        pages = {1990}
}

@ARTICLE{Nelson2019b,
       author = {{Nelson}, Dylan and {\noopsort{b}}{Pillepich}, Annalisa and {Springel}, Volker and {Pakmor}, R{\"u}diger and {Weinberger}, Rainer and {Genel}, Shy and {Torrey}, Paul and {Vogelsberger}, Mark and {Marinacci}, Federico and {Hernquist}, Lars},
        title = "{First results from the TNG50 simulation: galactic outflows driven by supernovae and black hole feedback}",
      journal = {\mnras},
         year = 2019,
        month = dec,
       volume = {490},
       number = {3},
        pages = {3234-3261},
          doi = {10.1093/mnras/stz2306}
}

@ARTICLE{Pillepich2019,
       author = {{Pillepich}, Annalisa and {Nelson}, Dylan and {Springel}, Volker and {Pakmor}, R{\"u}diger and {Torrey}, Paul and {Weinberger}, Rainer and {Vogelsberger}, Mark and {Marinacci}, Federico and {Genel}, Shy and {van der Wel}, Arjen and {Hernquist}, Lars},
        title = "{First results from the TNG50 simulation: the evolution of stellar and gaseous discs across cosmic time}",
      journal = {\mnras},
         year = 2019,
        month = dec,
       volume = {490},
       number = {3},
        pages = {3196-3233},
          doi = {10.1093/mnras/stz2338}
}

@ARTICLE{RodriguezGomez2019,
       author = {{Rodriguez-Gomez}, V. and {Snyder}, G.~F. and {Lotz}, J.~M. and {Nelson}, D. and {Pillepich}, A. and {Springel}, V. and {Genel}, S. and {Weinberger}, R. and {Tacchella}, S. and {Pakmor}, R. and {Torrey}, P. and {Marinacci}, F. and {Vogelsberger}, M. and {Hernquist}, L. and {Thilker}, D.~A.},
        title = "{The optical morphologies of galaxies in the IllustrisTNG simulation: a comparison to Pan-STARRS observations}",
      journal = {\mnras},
         year = 2019,
        month = mar,
       volume = 483,
        pages = {4140-4159},
          doi = {10.1093/mnras/sty3345}
}

@ARTICLE{Schulz2020,
       author = {{Schulz}, Sebastian and {Popping}, Gerg{\"o} and {Pillepich}, Annalisa and {Nelson}, Dylan and {Vogelsberger}, Mark and {Marinacci}, Federico and {Hernquist}, Lars},
        title = "{A redshift-dependent IRX-{\ensuremath{\beta}} dust attenuation relation for TNG50 galaxies}",
      journal = {\mnras},
         year = 2020,
        month = oct,
       volume = {497},
       number = {4},
        pages = {4773-4794},
          doi = {10.1093/mnras/staa1900}
}

@ARTICLE{Steinacker2013,
       author = {{Steinacker}, J{\"u}rgen and {Baes}, Maarten and {Gordon}, Karl D.},
        title = "{Three-Dimensional Dust Radiative Transfer}",
      journal = {\araa},
         year = 2013,
        month = aug,
       volume = {51},
       number = {1},
        pages = {63-104},
          doi = {10.1146/annurev-astro-082812-141042}
}

@ARTICLE{Torrey2015,
       author = {{Torrey}, Paul and {Snyder}, Gregory F. and {Vogelsberger}, Mark and {Hayward}, Christopher C. and {Genel}, Shy and {Sijacki}, Debora and {Springel}, Volker and {Hernquist}, Lars and {Nelson}, Dylan and {Kriek}, Mariska and {Pillepich}, Annalisa and {Sales}, Laura V. and {McBride}, Cameron K.},
        title = "{Synthetic galaxy images and spectra from the Illustris simulation}",
      journal = {\mnras},
         year = 2015,
        month = mar,
       volume = {447},
       number = {3},
        pages = {2753-2771},
          doi = {10.1093/mnras/stu2592}
}

@ARTICLE{Tortora2024,
       author = {{Tortora}, C. and {Ragusa}, R. and {Gatto}, M. and {Spavone}, M. and {Hunt}, L. and {Ripepi}, V. and {Dall'Ora}, M. and {Abdurro'uf} and {Annibali}, F. and {Baes}, M. and {Belfiore}, F.~M.~C. and {Bellucco}, N. and {Bolzonella}, M. and {Cantiello}, M. and {Dimauro}, P. and {Kluge}, M. and {Lelli}, F. and {Napolitano}, N.~R. and {Nucita}, A. and {Radovich}, M. and {Scaramella}, R. and {Schinnerer}, E. and {Testa}, V. and {Unni}, A.},
        title = "{VST-SMASH: the VST Survey of Mass Assembly and Structural Hierarchy}",
      journal = {The Messenger},
         year = 2024,
        month = sep,
       volume = {193},
        pages = {31-34},
          doi = {10.18727/0722-6691/5366}
}

@ARTICLE{Trayford2017,
       author = {{Trayford}, James W. and {Camps}, Peter and {Theuns}, Tom and {Baes}, Maarten and {Bower}, Richard G. and {Crain}, Robert A. and {Gunawardhana}, Madusha L.~P. and {Schaller}, Matthieu and {Schaye}, Joop and {Frenk}, Carlos S.},
        title = "{Optical colours and spectral indices of z = 0.1 eagle galaxies with the 3D dust radiative transfer code skirt}",
      journal = {\mnras},
         year = 2017,
        month = sep,
       volume = {470},
       number = {1},
        pages = {771-799},
          doi = {10.1093/mnras/stx1051}
}
\bibliographystyle{aasjournalv7}


\end{document}